\documentclass[twocolumn,showpacs,preprintnumbers,amsmath,amssymb]{revtex4}

\usepackage{epsfig}
\usepackage{dcolumn}
\usepackage{bm}

\begin{document}

\title{Mach-Zehnder interferometer in the Fractional
Quantum Hall regime}

\author{Vadim V. Ponomarenko\footnote{also at St. Petersburg State
Polytechnical University, Center for Advanced Studies,
St. Petersburg 195251, Russia.}}
\affiliation{International Center for Condensed Matter
Physics, Universidade de Brasilia, 70904 Brasilia,
Brazil}
\author{Dmitri V. Averin}
\affiliation{Department of Physics and Astronomy,
University of Stony Brook, SUNY, Stony Brook, NY 11794}

\date{\today}


\begin{abstract}
We consider tunneling between two edges of Quantum Hall liquids
(QHL) of filling factors $\nu_{0,1}=1/(2 m_{0,1}+1)$, with $m_0 \geq
m_1\geq 0$, through two point contacts forming Mach-Zehnder
interferometer. Quasiparticle description of the interferometer is
derived explicitly through the instanton duality transformation of
the initial electron model. For $m_{0}+m_{1}+1\equiv m>1$, tunneling
of quasiparticles of charge $e/m$ leads to non-trivial $m$-state
dynamics of effective flux through the interferometer, which
restores the regular ``electron'' periodicity of the current in
flux. The exact solution available for equal propagation times
between the contacts of interferometer shows that the interference
pattern depends in this case on voltage and temperature only through
a common amplitude.
\end{abstract}

\pacs{73.43.Jn, 71.10.Pm, 73.23.Ad}

\maketitle

An electronic Mach-Zehnder interferometer (MZI) realized recently
\cite{mz1} at Weizmann institute consists of the two point contacts
between two single-mode edges of the Integer Quantum Hall liquids.
Its transport properties exhibit strongly pronounced electron
interference sensitive in experiments to charging effects. MZI in
the regime of the Fractional Quantum Hall effect (FQHE)
\cite{mz2,mz3} and more complicated structures including it
\cite{mz4} were studied theoretically in search for signatures of
the fractional statistics of FQHE quasiparticles. Some of these
theories, however, (cf. \cite{mz2} and \cite{mz3}) were based on
different postulated models of the quasiparticle transport in MZI
and obtained conflicting result, e.g., different periods of the
tunnel current modulation by external magnetic flux $\Phi_{ex}$
through the interferometer. The goal of this work is to develop the
theory of symmetric MZI in the FQHE regime that is valid for
arbitrary tunneling strength in its point contacts. In this theory,
quasiparticles are derived consistently from the standard model of
electron tunneling in the weak-tunneling limit. Scaling of electron
tunneling amplitudes up with voltage or temperature to the
strong-tunneling limit (similar to that in one point contact
\cite{kf}) generates non-trivial model of quasiparticle tunneling in
MZI as a dual to weak electron tunneling.

The main qualitative elements of our approach can be summarized as
follows. The phase difference between the two point contacts
expressed in terms of the effective flux $\Phi$ through the MZI
contains, in addition to the external flux $\Phi_{ex}$, a
statistical contribution. This contribution emerges, since each
electron coherently tunneling at different contacts changes $\Phi$
by $\pm m\Phi_0$, where $m=m_0+m_1+1$ and $\Phi_0$ is the flux
quantum. As a result, the system has $m$ quantum states which differ
by number of flux quanta modulo $m$ which are not mixed by
perturbative electron tunneling. However, in the non-perturbative
regime of strong tunneling, the states are mixed as $\Phi$ is
changed by $\pm \Phi_0$ by tunneling of individual quasiparticles.
This implies that the quasiparticles have to carry the charge $e/m$
and coincide with the quasiparticles
$e^*=2e\nu_0\nu_1/(\nu_0+\nu_1)=e/m$ in one point contact \cite{cf}.
(Flux dynamics in the MZI is similar to that in the antidot
tunneling \cite{ant}, where, however, $m=m_0-m_1$, and the $e/m$
quasiparticles are different from those in individual contacts.) The
quasiparticle Lagrangian for MZI derived below is a mathematical
expression of the flux-induced electron-quasiparticle transmutation.
If the times $t_0$ and $t_1$ of propagation between the contacts
along the two edges of the interferometer are equal: $\Delta t
\equiv t_0-t_1 =0$, the quasiparticle Lagrangian can be solved by
methods of the exactly solvable models. The resultant expression for
the tunneling current shows the crossover from the quasiparticle
tunneling at large voltages to the electron tunneling at low
voltages. Our results correct Ref.~\onlinecite{mz2} by showing that
the quasiparticle model used in that work does not correspond in the
weak-tunneling limit to electron tunneling at two separate point
contacts, and also restrict the validity of the quasiparticle
current found in \cite{mz3} to the leading term in the large-$V$
asymptotics.

In details, we start with the electronic model of MZI (Fig.~1)
formed by two single-mode edges with filling factors $\nu_l=1/
(2m_l+1)$, $l=0,1$, which differ from the model studied in
\cite{ant} only by the direction of propagation of one edge.
\begin{figure}[htb]
\setlength{\unitlength}{1.0in}
\begin{picture}(2.3,1.16)
\put(0.16,-0.15){\epsfxsize=2.00in \epsfbox{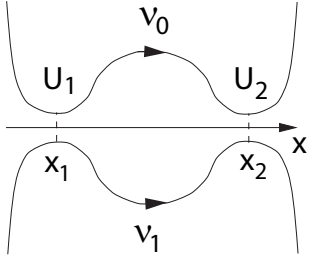}}
\end{picture}
\caption{Mach-Zehnder interferometer considered in this work: two
point contacts with tunneling amplitudes $U_j$ formed at points
$x_j$, $j=1,2$, between two single-mode edges with filling factors
$\nu_0$ and $\nu_1$ propagating in the same direction.}
\end{figure}
In the standard bosonization approach \cite{wen},
Lagrangian of weak electron tunneling in the two
contacts \cite{mz3} can be expressed through two bosonic
fields $\phi_l$ related to the correspondent edge
density $\rho_l(x,\tau)= (\sqrt{\nu_l}/2 \pi)
\partial_x \phi_l(x, \tau)$ as follows:
\begin{eqnarray}
{\cal L}_t\!\! =\!\! \sum_{j=1,2}& & \!\!\!\!
[\frac{DU_j}{2\pi} e^{i\kappa_j} e^{i \lambda \varphi_j}
+ h.c.] \equiv\!\!
\sum_{j=1,2} (T_j^+ +T_j^-) \label{e2}\\
 \lambda \varphi_j(t)\!\! &\equiv&\!\!
\frac{\phi_0(x_j,t)}{\sqrt{\nu_0}} -
\frac{\phi_1(x_j,t)}{ \sqrt{\nu_1} } \, , \;\;\;
\lambda=\sqrt{2m} \ , \label{e2a}
\end{eqnarray}
where $D$ is a common energy cut-off of the edge modes,
$U_j$ and $\kappa_j$ are the absolute values and the
phases of the dimensionless tunneling amplitudes.
Dynamics of the operators $\varphi_j$ is defined by the
Fourier transform of the imaginary-time-ordered
correlators $\langle \phi_l (x) \phi_p \rangle_\omega =
\delta_{lp} g(x/v_l,\omega)$, where (see, e.g.,
\cite{we}):
\begin{equation} g(z,\omega)=\frac{2 \pi}{\omega} \mbox{sgn} (z)
\Big(-\frac{1}{2}+\theta(\omega z)e^{-\omega z} \Big) \, ,
\label{e1}
\end{equation} and the
first term in brackets gives the usual equal-time commutation
relations $[\phi_l(x),\phi_p(0)]=i \pi \mbox{sgn}(x)\, \delta_{lp}$.
The phases $\kappa_j$ include contributions from the external
magnetic flux $\Phi_{ex}$ and from the average electron numbers
$N_{0,1}$ on the two sides of the interferometer: $\kappa_2 -
\kappa_1=2\pi[(\Phi_{ex}/\Phi_0)+ (N_0/\nu_0)-(N_1/\nu_1)]+
\mbox{const} \equiv \kappa $.

If a bias voltage $V$ is applied to the junction, the
operator of electron current from the edge $0$ into the
edge $1$ is $I^{e}=i\sum_{j=1,2}\sum_{\pm} (\pm)
T_j^{\pm} e^{\mp iVt}$. Its average contains the
phase-insensitive part $\bar{I}^e$ and the
phase-sensitive interference term $\Delta I^e(\kappa)$:
$I=\langle I^e \rangle =\bar{I}^e+\Delta I^e(\kappa)$.
At finite $V$, the phase difference $\kappa$ acquires
additional contribution related to $V$. For instance, if
the voltage changes only the electrochemical potential
of the edge $0$, and for perfect screening by external
gates, the phase varies as $\kappa(V)=\kappa+Vt_0$. In
the lowest non-vanishing order in $U_j$, the
phase-insensitive current consists of two contributions
from the individual point contacts $\bar{I}^e=\sum_j
\bar{I}^e_j$, which at temperature $T$ are \cite{kf}:
\begin{equation}
\bar{I}^e_j= (U^2_jD/2\pi) (2\pi T/D)^{\lambda^2-1}
C_{\lambda^2}(V/2\pi T)\, , \label{e3}
\end{equation}
where $C_g (v) \equiv \sinh (\pi v)|\Gamma (g/2 +iv)|^2/
[\pi\Gamma(g)]$ and, for $g$ equal to an even positive integer,
reduces to the polynomial $C_g (v)=v
\prod_{n=1}^{g/2-1}(n^2+v^2)/\Gamma(g)$ . The interference current
can be written as
\[ \Delta I^e= (\frac{U_1U_2D}{ \pi^2}) (\frac{\pi T}{D})^{\lambda^2
-1} \mbox{Im} \Big\{ \int^\infty_{-\infty} ds \sin
(\kappa(V)-V\bar{t}- \frac{sV}{\pi T}) \] \vspace{-3ex}
\begin{equation}
\cdot \prod_{l=0,1} [i\sinh(s-(-1)^l \Delta t\pi
T-i0)]^{-1/\nu_l} \Big\} \, . \label{e4}
\end{equation}
in the notation $t_{0,1}=\bar{t} \pm \Delta  t$. This expression
coincides (up to redefinition of the phase
$\kappa_V=\kappa(V)-V\bar{t}$) with the interference current in the
antidot geometry \cite{ant,gl}. One can evaluate the integral
(\ref{e4}) for integer $1/\nu_l$ by residues, and find the
visibility $\mbox{Vis} \equiv ( \max_\kappa I-\min_\kappa
I)/(\max_\kappa I+\min_\kappa I)$. For instance, for
$\nu_0=\nu_1\equiv \nu$ in the asymptotic regime $V \Delta t \gg 1$
and low temperatures, the visibility decreases with voltage
oscillating as
\begin{equation}
\mbox{Vis} \simeq{(2/\nu_0-1)! \over (1/\nu_0-1)! } {8
U_1U_2 \over U^2_1+U^2_2} {|\sin(V\Delta t)|\over
|2\Delta t V|^{1/\nu_1}}\, . \label{e7}
\end{equation}

In the opposite limit of $V,T< 1/\Delta t$, the integral
in Eq.\ (\ref{e4}) reduces to the same polynomial
$C_{\lambda^2}(V/2\pi T)$ as in Eq.\ (\ref{e3}), and the
full current $\langle I^e \rangle$ is specified by the
coherent sum of the two point-contact amplitudes:
\begin{equation}
I= |U_1+U_2e^{i\kappa_V }|^2T (2\pi T/D)^{\lambda^2-2}
C_{\lambda^2}(V/2\pi T)\, .\label{e9}
\end{equation}
In this regime, the visibility reaches its maximum
$2U_1U_2/(U_1^2+U_2^2)$. Naively, Eq.\ (\ref{e9}) seems to suggest
that for small $\Delta t$ the two-point-contact model of MZI reduces
to one point contact with the new tunnel amplitude. This, however,
is true only for weak tunneling. At large voltages or temperatures,
the system automatically flows into the regime of strong tunneling,
in which the model of two FQHLs strongly coupled at two separate
point contacts possesses non-trivial topology of quasiparticle
tunneling trajectories. Since FQHL is a topological quantum liquid
\cite{wen}, the non-trivial topology of the model implies multiple
degeneracy of its ground state which is absent in one point contact.

To derive the dual strong-coupling model for the MZI, we treat the
problem in imaginary time and follow a standard instanton technique.
The ground states are determined by minimization of the action
${\cal S}$ that consists of the tunneling part ${\cal S}_{t}$
(\ref{e2}) and kinetic term ${\cal S}_{kin}$ defined by the
correlators (\ref{e1}). In the limit $U_j\gg 1$, ${\cal S}_{t}$
gives the dominant contribution to the action, and it would be
natural to fix both tunneling modes $\varphi_j$ at the extrema of
their corresponding parts of Eq.~(\ref{e2}). These modes, however,
do not commute, $[\varphi_2,\varphi_1]=i \pi $, the fact reflected
in the interchange relation for the transfer terms (\ref{e2}) at the
two contacts:
\begin{equation}
T_2^{\pm}T_1^{\mp}=e^{2\pi m i} T_1^{\mp} T_2^{\pm} \, . \label{e10}
\end{equation}
We see that while different $T_j^{\pm}$ commute, each
interchange of electron transfers at the two contacts
adds statistical contribution $\pm m\Phi_0$ to the
external magnetic flux $\Phi_{ex}$ modifying the
interference phase $\kappa$. This flux dynamics affects
the perturbative expansion of the partition function in
$U_{1,2}$ by changing the phase branch of terms in the
expansion according to Eq. (\ref{e10}), when the
imaginary times of the transfer operators, $T_{1}^{\pm}$
and $T_{2}^{\mp}$, change order. In general, we can make
different choices of the phase branches multiplying the
operators $T_{j}^{\pm}$ by some Klein factors $\exp\{
\pm i\sqrt{2\gamma} \eta_j\}$ with arbitrary integers
$\gamma$ and the free zero-energy bosonic modes $\eta_j$
defined by their imaginary-time-ordered correlators:
$<T_\tau{\eta_i(\tau) \eta_j(0)}>=i \pi
\Theta((j-i)\tau) (1-\delta_{ij})$. For any $\gamma$,
incorporation of these Klein factors into $T_j^{\pm}$
(\ref{e2}) does not change the perturbation expansion of
the partition function in ${\cal S}_t$ in any order.
However, it affects the kinetic part of the action and
hence the ground-state energy. Indeed, the new tunneling
fields $\Phi_j=\lambda \varphi_j+ \sqrt{2\gamma} \eta_j$
are characterized by the kinetic action
\begin{eqnarray}
{\cal S}_{kin}&=&{1 \over 2} \int {d \omega \over 2\pi}
\sum_{i,j}\Phi_i(-\omega) \hat{K}_{i,j}^{-1}(\omega)
\Phi_j(\omega)\, ,
\label{e11} \\
\hat{K}(\omega)&=&\lambda^2 g(0,\omega) \hat{1}+ \sum_\pm [\mp {2
\pi \gamma \over \omega} +\sum_j {1\over \nu_j}g(\mp t_j,\omega)]
\hat{\sigma}_\pm \ , \ \nonumber
\end{eqnarray}
where $\hat{\sigma}_\pm$ are the raising and lowering $2 \times 2$
matrices. For well-separated contacts, $t_{0,1}D \gg 1$,
minimization of energy under the strong-tunneling conditions:
\begin{equation}
\Phi_j=2 \pi n_j-\kappa_j\equiv \Phi_{n_j}, \label{e12}
\end{equation}
gives $\gamma=m$ \cite{us}, the choice that also guarantees the
commutativity of the tunneling fields $\Phi_j$.

The standard instanton expansion of the partition function ${\cal
Z}$ for the degenerate ground states $(\Phi_{n_1}, \Phi_{n_2})$
(\ref{e12}) expresses it as ${\cal Z}=\sum_{n_j} {\cal
Z}_{n_1,n_2}$. Substitution of the asymptotic form of the expansion
around the state $(\Phi_{n_1}, \Phi_{n_2})$:
$\Phi_j(\tau)=\Phi_{n_j}+\sum_l 2\pi e_{l,j}
\theta(\tau-\tau_{l,j})$, into $\exp\{-{\cal S}(\Phi_1,\Phi_2)\}$,
followed by summation over the numbers of instantons/anti-instantons
$e_{l,j}=\pm 1$ and integration over their times $\tau_{l,j}$ gives:
\[ {\cal Z}_{n_1,n_2}\propto \int\!\!D\Theta_{1,2} \exp\{-{\cal
S}^D_{kin}+ \]

\vspace{-3ex}

\begin{equation}
 \sum_j\!{W_j D\over 2 \pi}\!\!\int\!\!d\tau \cos
[\Theta_j(\tau) +{\kappa_j-2 \pi n_j \over (-1)^jm}] \} \label{e14}
\end{equation} with a constant of proportionality independent of
$n_{1,2}$, and
\begin{equation}
{\cal S}^D_{kin}(\Theta)={1 \over 2} \int {d \omega \over 2\pi}
\Theta(-\omega) [({2\pi\over \omega})^2\hat{K}^{-1}(\omega)]^{-1}
\Theta(\omega). \label{e15}
\end{equation}
Comparing the  $\Theta$-correlators defined by action (\ref{e15})
with $g(z,\omega)$ (\ref{e1}), we separate these fields into two
parts: $\Theta_j=(-1)^j [(2/m)^{1/2} \eta_j+ (2/ \lambda)
\vartheta_{j}]$, with statistical terms $\eta_{1,2}$, and the chiral
fields $\vartheta$:
\begin{equation}
<\vartheta_{j}^2>=g(0,\omega)\,,\ <\vartheta_{2}
\vartheta_{1}>= {g(t_0,\omega)\over
\nu_0\lambda^2}+{g(t_1,\omega)\over \nu_1\lambda^2}\,.
\label{e17}
\end{equation}

Since the terms ${\cal Z}_{n_1,n_2}$ (\ref{e14}) depend on $n_{1,2}$
only through their difference modulo $m$, the partition function
${\cal Z}$ becomes a finite sum up to an irrelevant (but divergent)
factor. Summation over $n_1-n_2$ combined with integration over the
new statistical fields can be reduced then to the trace over an
$m$-dimensional Hilbert space, if a proper $m$-dimensional matrix is
ascribed to each instanton tunneling exponent in (\ref{e14}). These
unitary matrices $\bar{F}_j$ are characterized by the following
relations:
\begin{equation}
\bar{F}_1\bar{F}_2=e^{2 \pi i\over m} \bar{F}_2\bar{F}_1
, \;\;\; \langle \bar{F}^k_1 (\bar{F}^+_1)^p\bar{F}^l_2
(\bar{F}^+_2)^q \rangle =\delta_{kp}\delta_{lq}\, ,
\label{e18}
\end{equation}
where the Kronecker symbol $\delta_{ij}$ is defined modulo $m$. The
first relation in (\ref{e18}) is due to the statistical parts of the
fields $\Theta_j$, while the second one follows from the
$m$-periodic dependence of (\ref{e14}) on $n_{1,2}$. Writing ${\cal
Z}$ as a trace makes possible to interpret it as a partition
function of quasiparticles with tunneling Lagrangian $\bar{{\cal
L}}_t$ of the real-time form dual to the Lagrangian (\ref{e2}):
\[ \bar{{\cal L}}_t=\sum_{j=1,2}\Big[ {W_jD\over 2\pi } \bar{F}_j
\exp \big\{ i \big({\kappa_j(V) \over m} +{2\vartheta_j \over
\lambda} - {Vt \over m}\big)\big\}  + h.c.\Big]
\]

\vspace{-3 ex}

\begin{equation}
\equiv \sum_{j=1,2} \sum_{\pm} \bar{T}_j^{\pm} e^{\mp
iVt /m} \, . \label{e19} \end{equation} The operators
$\bar{F}_j$ are the quasiparticle Klein factors
describing their statistics and acting in the space of
the $m$-degenerate (in the absence of quasiparticle
tunneling) ground state of the MZI. The current
associated with the quasiparticle tunneling is expressed
as usual in terms of the transfer operators: $I^q=(i/m)
\sum_{j=1,2}\sum_{\pm} \pm \bar{T}_j^{\pm} e^{\mp
iVt/m}$. The model (\ref{e19}) with the Klein factors
(\ref{e18}) for quasiparticle tunneling between the
edges of the MZI is a direct analogue of the
quasiparticle model we derived earlier for the antidot
\cite{ant}. At $\nu_0 = \nu_1$ it coincides with the
postulated quasiparticle model of \cite{mz3} for a
special choice of matrices $\bar{F}_j$.

In the case of symmetric interferometer, with $\Delta
t=0$ and equal velocities of the edge modes $\phi_l$,
both tunneling operators $\varphi_j$ in Eq. (\ref{e2})
are the operator values of the chiral bosonic field
$\phi_-$ at two points $x_{1,2}$. The field $\phi_-$ is
composed of the incoming edge modes $\phi_{0,1}$ as in
Eq.~(\ref{e2a}). Strong tunneling at the point contacts
can be described by imposing the Dirichlet boundary
condition on $\phi_-$, the ``unfolded'' form of which
\cite{bc} implies a free chiral propagation of the
fields $\mbox{sgn}(x-x_j)(\phi_-(x)+\kappa_j/\lambda)$
across the point $x_j$. Deviation from their free
propagation is driven by the  dual tunneling terms
$\bar{\cal L}_{t,j}=D W_j \cos(2/\lambda
(\phi_-(x_j)+\kappa_j/\lambda ))/\pi$. Successive
application of these boundary conditions at the two
contacts gives free propagation of the dual chiral field
\begin{eqnarray}
\vartheta_-(x)\!\!&=&\!\!\phi_-(x)\theta(x_1-x)+(\phi_-(x)-2 {\kappa
\over \lambda})\theta(x-x_2) \nonumber \\
&-&\!\!(\phi_-(x)+2 {\kappa_1 \over \lambda})\theta(x-x_1)
\theta(x_2-x)\, . \label{e20}
\end{eqnarray}
Substitution of this field into $\bar{\cal L}_{t,j}$ and
further comparison of the result with Eqs. (\ref{e17})
and (\ref{e19}) prove that
$\vartheta_j=\vartheta_-(x_j)$. Applied voltage changes
$\vartheta_-$ into $\vartheta_--Vt/\lambda$.

The derived quasiparticle model (\ref{e19}) is exactly
solvable at $\lambda=2$ (i.e., when $\nu_0=1/3$ and
$\nu_1=1$) by fermionization. Indeed, the Klein factors
for $m=2$ can be represented by two Pauli matrices and
fermionized as $\bar{F}_j=i \xi_j \xi_0$ in terms of
three Majorana fermions $\{\xi_n,\xi_{n'}\}_+=2
\delta_{n,n'}$. Introducing a chiral fermion field as
$\psi=\xi_0 \sqrt{D/(2 \pi v)}\exp(i \vartheta_-)$ we
come to the Hamiltonian
\begin{equation}
{\cal H}=-i v\! \int\! dx\, \psi^+
\partial_x \psi- v\sum_j [w_j \xi_j \psi^+(x_j)+h.c.] \, ,
 \label{e21}
\end{equation}

\vspace{-4ex}

\[ w_j= i(D/2 \pi v)^{1/2} W_j e^{-{i\over 2}\kappa_{j,V}} , \]
where the applied voltage is accounted for by the fermion chemical
potential equal to $V/2$. Note that the Hamiltonian (\ref{e21}), in
contrast to the fermionic Hamiltonian of \cite{mz2}, contains two
different Majorana fermions at the two tunneling points $x_{1,2}$.
As a result, the Heisenberg equation of motion describes scattering
at the two points of the field $\psi(x,t)$, elsewhere exhibiting a
free chiral propagation, with two \emph{disentangled} matching
conditions
\begin{equation}
\! i\psi(x)|^{x_j+0}_{x_j-0} \! = \! w_j \xi_j , \;
 \partial_t\xi_j(t)\! = \! 2iv[w_j\psi^+(x_j,t)-h.c.] \, .
\label{e22} \end{equation} Away from the tunneling points, the field
$\psi(x,t)$ propagates freely and can be represented as
$\psi(x,t)=\int dk \psi_k \exp\{ik(x-vt)\}/(2 \pi)$. Solution of
each of the condition (\ref{e22}) defines then a $(2 \times 2)$
scattering matrix $\hat{\cal S}_{j,k}$ of particle and hole
$(\psi_k,\psi^+_{-k})$ at contact $j$:
\begin{equation}
{\cal S}^{\pm \pm}_{j,k}={k \over k+i2 |w_j|^2}\, , \;\; {\cal S}^{-
+}_{j,k}={2iw_j^2\over k+i2 |w_j|^2}\, . \label{e23}
\end{equation}
Successive particle-hole scattering at the two points is governed by
the scattering matrix $\hat{\cal S}_k=\hat{\cal S}_{2,k}\hat{\cal
S}_{1,k}$ which determines the average tunneling current:
\begin{equation}
I=\int_0^{V \over 2v} {v dk \over2 \pi}{|2ik \sum_j
w_j^2|^2\over\prod|(k+2i|w_l|^2)|^2} \, . \label{e24}
\end{equation}
Introducing $\Gamma_j=DW_j^2/\pi$, one expresses the current as
\begin{equation}
I={|\Gamma_1e^{i\kappa_V}+\Gamma_2|^2 \over \Gamma_1^2 -\Gamma_2^2}
[I_{1/2}(V,\Gamma_2)-I_{1/2}(V,\Gamma_1)]\, , \label{e25}
\end{equation}
where  $I_{1/2}(V,\Gamma)$ is the tunneling current in a single
point contact \cite{kf}: $I_{1/2}(V,\Gamma)= V/(4\pi)
-\Gamma/(2\pi)\arctan(V/2\Gamma)$. The low voltage asymptotics of
$I$ is proportional to $V^3$ and coincides with the electron
tunneling current in Eq.~(\ref{e9}) under the condition $U_j=\pi
W_j^{-2}/2$ expected from the single-point-contact duality. Equation
(\ref{e25}) holds at finite temperature $T$.

This calculation of the current can be generalized to
other values of $\lambda^2=2m$, for which a
thermodynamic Bethe ansatz solution is known \cite{ba}
for a single point contact. The solution exploits a set
of quasiparticle states describing $\vartheta_-(x)$
excitations and introduced through the massless limit of
a sine-Gordon model. These quasiparticles are kinks,
antikinks, and breathers of height defined by the
sine-Gordon interaction and equal to $\pi \lambda$. They
remain interacting in the massless limit as described by
a bulk S-matrix, but undergo separate one by one
scattering on the point contact specified with a one
particle boundary S-matrix. Their scattering on two
point contacts occurs successively and separately at
different points as follows from chiral dynamics of the
local $\vartheta_-(x)$ fluctuations at the point
contacts derived above through application of "unfolded"
Dirichlet boundary conditions. Therefore it is described
with product of two boundary S-matrices dependent on
$\kappa_{1,2}$, respectively. To obtain these matrices
from the one found in \cite{ba} in the case of
$\kappa=0$, we notice that each phase $\kappa_j$ in Eq.
(\ref{e18}) results from the shift of $\vartheta_-$ by
the constant $\kappa_j/\lambda$. Hence the operators
$\exp(\pm i\lambda \vartheta_-/2)$ of the $\vartheta_-$
kinks/antikinks acquire just constant phase factors
$e^{\pm i\kappa_j /2}$. The boundary S-matrix of
\cite{ba} transforms into
\begin{equation}
{\cal S}^{\pm \pm}_{j,k}={(k/T_{jB})^{m-1}
e^{i\alpha_k}\over 1+i(k/T_{jB})^{m-1} }\, , {\cal S}^{-
+}_{j,k}={e^{i(\alpha_k-\kappa_{jV})}\over
1+i(k/T_{jB})^{m-1} }\, , \nonumber
\end{equation}
and the tunneling current produced by the kink-antikink
transitions breaking charge conservation takes for the
two point contact the following form
\begin{equation}
I=\int_0^{\infty} v dk |(\hat{\cal S}_2\hat{\cal
S}_1)^{-,+}|^2 n[f_+-f_-] \, .\label{e26}
\end{equation}
Notice that both, the density of states $n(k,V)$ and the
distribution functions $f_\pm$ for kinks and antikinks,
are defined by the "bulk" of the system and do not
depend on the scattering on impurities.  Then the
tunneling current in (\ref{e26}) takes the form that
generalizes Eq. (\ref{e25})
\begin{equation}
I={|T_{1B}^{m-1}e^{i\kappa_V}+T_{2B}^{m-1}|^2 \over
T_{1B}^{2(m-1)} -T_{2B}^{2(m-1)}}
(I_{1/m}(V,T_{2B})-I_{1/m}(V,T_{1B}))\, , \label{e28}
\end{equation}
where $I_{1/m}(V,T_{jB})$ is the tunneling current
through a single point contact between two effective
edges of the filling factor $1/m$. The energy scales
$T_{jB}$ are related to both correspondent electron and
quasiparticle tunneling amplitudes $U_j, W_j$ in the
same way as in the case of the individual point contact
\cite{ba}. This matches the low voltage dependence of
the current in (\ref{e28}) with the electron tunneling
current asymptotics in Eq. (\ref{e9}). Meanwhile, its
high voltage dependence asymptotically coincides with
the quasiparticle calculation in \cite{mz3}. Finally,
the expression (\ref{e28}) for the current shows that
the visibility of Eq. (\ref{e9}) does not vary with
temperature and voltage, while $V\Delta t,T\Delta t\ll
1$. The interference dependence of the current on the
external magnetic flux has the same form of a simple one
mode modulation, which is not affected by the change of
regimes from electron to quasiparticle tunneling.

In conclusion, we have derived the quasiparticle model
[Eqs.~(\ref{e17}), (\ref{e18}), and (\ref{e19})] of the
Mach-Zender interferometer in the FQHE regime for
arbitrary filling factors of interferometer edges of MZI
from its electron tunneling model. In the limit $\Delta
t=0$, this model allows an exact solution, which
describes the crossover from electron to quasiparticle
tunneling and shows that the interference pattern
remains the same in both regimes and is independent of
voltage and temperature.

This work was supported by the MCT of Brazil and the US
ARO under grant DAAD19-03-1-0126.

\end{document}